\documentclass[12pt]{article}
\usepackage{amssymb}

\newcommand{\bo}{{\bar o}}

\def\bo{{\raise.15ex\hbox{\large$\Box$}}}               

\def\face{{\raise.2ex\hbox{$\displaystyle \bigodot$}\mskip-2.2mu \llap {$\ddot
        \smile$}}}                                      

\def\leftrightarrowfill{$\mathsurround=0pt \mathord\leftarrow \mkern-6mu
        \cleaders\hbox{$\mkern-2mu \mathord- \mkern-2mu$}\hfill
        \mkern-6mu \mathord\rightarrow$}       
\def\dvec#1{\vbox{\ialign{##\crcr
        \leftrightarrowfill\crcr\noalign{\kern-1pt\nointerlineskip}
        $\hfil\displaystyle{#1}\hfil$\crcr}}}           

\def\beq{\begin{equation}}
\def\eeq{\end{equation}}

\def\beqx{\begin{displaymath}}
\def\eeqx{\end{displaymath}}

\def\beql{\begin{eqnarray}}
\def\eeql{\end{eqnarray}}

\newcommand{\bea}{\begin{eqnarray}}
\newcommand{\eea}{\end{eqnarray}}

\newcommand{\mod}{\;\;\;\;{\rm mod }\;}

\def\[{\left [}
\def\]{\right ]}
\def\({\left (}
\def\){\right )}

\def\+{\oplus}

\begin{document}

\hbox{\hskip 12cm NIKHEF/2005-018  \hfil}
\hbox{\hskip 12cm IMAFF/FM-05-21  \hfil}
\hbox{\hskip 12cm hep-th/0510074 \hfil}

\vskip .5in

\begin{center}
{\Large \bf Remarks on Global Anomalies in RCFT Orientifolds }

\vspace*{.7in}
{ B. Gato-Rivera}$^{a,b}$
{and A.N. Schellekens}$^{a,b,c}$
\\
\vskip .5in

${ }^a$ {\em NIKHEF Theory Group, Kruislaan 409, \\
1098 SJ Amsterdam, The Netherlands} \\

\vskip .2in

${ }^b$ {\em Instituto de Matem\'aticas y F\'\i sica Fundamental, CSIC, \\
Serrano 123, Madrid 28006, Spain} \\

\vskip .2in

${ }^c$ {\em IMAPP, Radboud Universiteit,  Nijmegen}

\end{center}

\begin{center}
\vspace*{0.5in}
{\bf Abstract}

\end{center}

We check the list of supersymmetric
standard model orientifold spectra of Dijkstra, Huiszoon and Schellekens
for the presence of global anomalies, using probe branes. Absence of global anomalies
is found to impose strong constraints, but in nearly all
cases they are automatically
satisfied by the solutions to the tadpole cancellation conditions.

\vskip 1in

\noindent
October 2005

\newpage

\section{Introduction}

In previous papers \cite{Dijkstra:2004ym}\cite{Dijkstra:2004cc} co-authored by one of us a large number
of supersymmetric open string spectra was found with a chiral
spectrum that exactly matches the standard model spectrum.
These models were constructed using orientifolds of tensor products
of N=2 minimal models. The standard model
gauge groups arise due to Chan-Paton multiplicities
of boundary states of the underlying rational conformal field theory.

In contrast to the majority of published work on orientifold model building 
(see {\it e.g.} \cite{Blumenhagen:2005mu} and references therein), the
construction of \cite{Dijkstra:2004cc} is algebraic and not geometric. It is based on
rational conformal field theory (RCFT) on surfaces with boundaries
and crosscaps. The basic RCFT building blocks and the way they are
put together are subject to a set of constraints which are the result
of many years of work by several groups.

The constraints can be divided into world-sheet and space-time
conditions.
The boundary states themselves must satisfy the ``sewing constraints" \cite{Lewellen:1991tb}
\cite{Cardy:1991tv}\cite{Sagnotti:1996eb}\cite{Pradisi:1996yd}\cite{Behrend:1999bn}.
There are further constraints on the crosscap states needed to define
non-orientable surfaces\cite{Fioravanti:1993hf}\cite{Pradisi:1995pp}. 
These are all worldsheet conditions needed to guarantee
the correct factorization of all amplitudes. In addition some
space-time conditions must be imposed, the tadpole cancellation conditions.
They are needed to make sure that tree-level one-point functions of closed string
states on the crosscap cancel those of the disk. If these tadpoles are left
uncancelled, this will manifest itself in the form of infinities in sum of the Klein
bottle, annulus and Moebius diagrams. 

If the tadpoles correspond to physical states in the projected closed
string spectrum, these infinities merely signal that
the corresponding string theory
is unstable and might be stabilized by shifting the vacuum expectation
value of the corresponding field.
However, if the tadpoles do not correspond to physical states 
their presence implies a fundamental inconsistency in the theory, which
may manifest itself in the form of
chiral anomalies in local gauge or gravitational
symmetries.

There is no proof that the aforementioned set of conditions is sufficient to
guarantee consistency of the resulting unoriented, open string theories. It was
shown in \cite{Fuchs:2004dz} that for the 
simple current boundary states derived in \cite{FHSSW} and that where used in \cite{Dijkstra:2004cc}
all sewing constraints
are satisfied in the oriented case. To our knowledge, however,
there is still no complete proof in the unoriented case, although
important progress was made in \cite{Fjelstad:2005ua}. Nevertheless, the
boundary and crosscap states used in the construction are based on generic
simple current modifications of the Cardy boundary states \cite{Cardy} and
the Rome crosscap formula \cite{PlanarDuality}. They have been succesfully compared with geometric
constructions, for example the circle and its orbifold \cite{Dijkstra:2003th} and WZW models. In addition,
they can be shown to yield integral partition functions in all cases, a
highly non-trivial requirement \cite{thesis_lennaert}.

With regard to space-time consistency there is a more concrete reason to worry. In
other constructions of orientifold models it was observed that in certain cases
gauge groups with global anomalies can occur, even though all tadpole conditions are
satisfied \cite{Uranga:2000xp}\cite{Marchesano:2004xz}\cite{Blumenhagen:2005tn}. 
By ``global anomalies" we mean here anomalies in the global definition of
field theory path integral, as first described in \cite{Witten:1982fp}. The symptom for such an
anomaly is an odd number of massless fermions in the vector representation of a 
symplectic factor of the gauge group (including in particular doublets of $SU(2)$).

Examples of orientifold spectra having such a problem were found in geometric
settings, where the problem can be traced back to uncancelled K-theory charges
of branes and O-planes. It is known that D-branes are not characterized by (co)homology
but by K-theory \cite{Witten:1998cd}\cite{Minasian:1997mm}. 
Tadpole cancellation guarantees in particular the cancellation of
cohomology charges of branes, which are characterized by long range RR fields coupling
to these charges. This cancellation is physically necessary for branes and O-planes
that fill all non-compact dimensions, since the field of an uncancelled charge
cannot escape to infinity.
However, the branes may carry additional $Z_2$-charges without
a corresponding long range field. Tadpole cancellation does not
imply the cancellation of these charges.
If they remain uncancelled, this may manifest
itself in the form of global anomalies.

This implies that also in algebraic constructions one has to be prepared
for the possibility of additional constraints. Unfortunately a complete description
of global anomalies in theories of unoriented open strings does not seem to be
available at present. Therefore the best we can do is to examine if the symptoms of the
problem are present.

This check was not done systematically for the results presented in \cite{Dijkstra:2004cc}. However,
since all gauge groups and representations were stored, we have been able to do an
{\it a posteriori} check. This leads to the following results. The total number of complete
spectra at our disposal is 270058\rlap.\footnote{This is larger than the number of spectra mentioned in
\cite{Dijkstra:2004cc} because the latter were obtained after identifying spectra modulo hidden sector
details. In other words, some of the 270058 stored spectra differ only in the hidden sector.}
Of the 270058 models, only 1015 turn out to have one or more globally anomalous symplectic
factors. Interestingly, on average the anomalous models have more than one anomalous
symplectic factor:
there are 2075 anomalous symplectic factors out of a total of 845513.

The gauge groups of these  models usually (but not always) have ``hidden sectors"
in addition to the standard model
gauge group $SU(3) \times SU(2) \times U(1)$. Since the standard model itself is free of global
anomalies, the origin of the anomaly is always related to the hidden sector, but this may
happen in two ways. First of all a symplectic factor within the hidden sector may be anomalous.
However, the hidden sector can also cause an $SU(2)$ factor (the weak
gauge group or, in a subclass of models, an additional $Sp(2)$ factor)
of the standard model to be anomalous.
Since we require any open string stretching between the standard model and hidden branes to be
non-chiral, this can only happen if an open string has one end on the
standard model $SU(2)$ and the other
end on a brane with an $O(N)$ Chan-Paton group, with $N$ odd (if the other end is on a symplectic
brane the ground state dimension is automatically even, and when it ends on a complex brane
the ground state must be a non-chiral pair, again yielding an even multiplicity).
This does indeed occur in a few of the aforementioned anomalous cases.

Even if the massless spectrum does not exhibit this problem, this still does not
guarantee that the corresponding string theory is globally consistent. 
Indeed, if one starts with string theory with a globally anomalous $Sp(2)$ factor, moving
the two symplectic branes away from the orientifold plane produces a $U(1)$ theory which
presumbably is also globally inconsistent,
 but which does not exhibit the problem in its
field theory limit. We assume here and below that continously moving branes cannot
introduce or remove such an inconsistency.

A more
powerful constraint was suggested in \cite{Uranga:2000xp}.  In addition to the CP-factors or branes
present in a given model, one may introduce ``probe-branes". The idea is to add a 
brane-antibrane pair to a given brane configuration, which we assume to have
a field theory limit without global anomalies (and that is tadpole free, and
hence has also no local anomalies).
The reason for adding such pair rather than a single brane is that
the brane and anti-brane cancel each others cohomological brane charges, and
hence one does not introduce couplings to long range RR fields. This implies
that the result is at least free of
local chiral anomalies. If that were not the case,
a discussion of global anomalies would
not make much sense.
The resulting configuration is not free of all tadpoles
(dilaton tadpoles will not cancel, for instance), and
in particular is neither supersymmetric nor stable,
but that should not affect the consistency.

The CP gauge group of the new configuration can now be checked for global anomalies.
Since the
probes are added in pairs, they cannot introduce new global field theory
anomalies in the existing configuration,
but if the CP groups of the probe-brane pair are symplectic,
one may find that the latter gauge groups have a global anomaly ({\it i.e.} an odd
number of vectors).
In that case one should conclude that the original theory was inconsistent as well.

It is not clear that this constraint captures all possible global string anomalies.
In the context of our RCFT construction, we have for every
choice of $N=2$ tensor product and modular invariant partition function
a definite number of distinct boundary
states at our disposal. For a given orientifold choice,
a certain subset of those boundaries
will have symplectic CP-factors. Each of them can (and will) be used, with its anti-brane,
as a probe brane pair.
If the algebraic model is viewed from a geometric point of view,
perhaps additional branes can be considered that do not have an algebraic description,
and that would lead to additional constraints if used as probes.
In all cases studied in the literature, the branes not present in the algebraic description
simply correspond to rational branes continuously moved to non-rational
positions. If this is also true for the more complicated
cases considered here, this would not yield anything new. The fact that the
set of RCFT boundaries is algebraically complete \cite{Pradisi:1996yd} may imply that they also
provide a complete set of probe brane constraints.

The boundary states considered so far all correspond geometrically to
space-time filling branes.
One may also use probe branes that
are not space-time filling, and indeed the corresponding
constraints are important to understand the
relation between tadpole cancellation and cancellation of local anomalies \cite{Uranga:2000xp}.
For example in non-chiral theories there may be unphysical tadpoles, but
they cannot manifest themselves as chiral anomalies. Instead they will then
appear as local anomalies of gauge theories on lower-dimensional branes.
However, we
expect lower dimensional branes to be irrelevant for global anomalies, because there
are no global anomalies in 1,2 and 3 dimensions. Nevertheless, clearly a
more fundamental discussion of these additional constraints in RCFT constructions
is needed, presumably involving the appropriate generalization of K-theory
charges to boundary and crosscap states.

In any case, the probe branes described above
imply, in general, a very large number of additional constraints. Typically, the models
we consider have a few thousand boundary states, and a few hundred of them yield a symplectic
gauge group. In principle, any
such gauge group imposes a mod-2 condition on the spectrum, and hence each
might reduce the number of solutions by a factor of 2.
Purely statistically speaking,
this could reduce the number of solutions by several orders
of magnitude. The aforementioned discussion of manifest global anomalies
suggests already that the result will be less dramatic.

The probe-brane constraint cannot be checked as easily as
the manifest global anomalies, because the probe
brane CP-factors are not listed in our database unless they happen
to be part of the CP gauge group themselves. The only way to check it
is to generate the models again, and re-compute the spectrum in the presence of any
probe branes that might occur. Even if a previously recorded spectrum has a global
inconsistency, it might still be possible to make a different choice for the hidden sector
gauge group and cancel it. However, rather surprisingly this was rarely necessary.
For the vast majority of MIPFs we did not encounter any global probe brane
anomalies for any solution, even though the number of potential anomalies was often
very large.
In \cite{Dijkstra:2004cc} tadpole solutions were obtained for 333 modular invariant 
partition functions.
We have encountered global anomalies for only 25 of those (and then only for
very few solutions in each case).
These MIPFs are listed in the table. We present this list because we hope 
that the presence of global anomalies
in these cases makes sense from another point of view, for example Calabi-Yau geometry.

It is not possible to state exactly which fraction of the 270058
stored spectra fails the probe brane conditions, because we do not have
sufficient information in some older cases to re-generate them exactly.
Instead, we
simply searched the corresponding MIPFs again with the probe brane condition imposed
as an additional constraint. In nearly all previous cases a new, global anomaly-free
solution turns out to exist. To get an idea of the effect of the probe brane
constraint on the original database, consider the MIPF that contributed the largest number of solutions,
the one listed in the table for tensor product $(1,6,46,46)$. For this MIPF
we had 19644 full tadpole solutions stored, including the precise
boundary labels needed to regenerate them (16243 of these 19644 solutions
are distinct if hidden sector details are ignored). Only 59 of the 19644
violated the global anomaly conditions, and in 8 of those 59 cases a new
solution was found that is free of global anomalies.

This is a very surprising result in view of the large number of constraints
implied by the probe brane procedure. In general, for every
boundary label $b$ with a symplectic CP group, one obtains a constraint
of the form
\beq \label{GlobalAnomaly}
 \sum_{i,a} N_a A^i_{ab} (\chi_i)_{0,L} = 0 \mod 2\ ,
\eeq
where $A^i_{ab}$ are the annulus coefficients and
$\chi_i$ is the Virasoro character of representation $i$ restricted
to massless characters of definite (in our case left, L) space-time chirality.
This imposes as many mod-2 constraints on the CP-multiplicities $N_a$
as there are symplectic boundaries.

This condition is similar but not identical to the chiral anomaly constraint
derived from tadpole cancellation
\beq \label{Anomaly}
\sum_{i,a} N_a w_i (A^i_{ab}  + 4 M^i_b) = 0
\eeq
where $M^i_{b}$ are the Moebius coefficients and $w_i$ is
the Witten index, $(\chi_i)_{0,L}-(\chi_i)_{0,R}$.
Here $b$ can be any boundary, not just those that appear in a given
solution with non-vanishing Chan-Paton multiplicity.
This implies the absence of local gauge anomalies associated
with probe branes. The index $b$ can represent any probe brane. The anti-branes
are not represented by any label in this set because they do not satisfy
the BPS condition for the given choice of unbroken supersymmetry. However,
we do know their anomaly contribution. Consider a $U(N)$ factor
in the configuration of interest, and a probe brane pair contributing
CP factors $U(M)_1 \times U(M)_2$ (consisting thus of four branes: a brane $b$, its conjugate,
$b^c$ and their anti-branes).
Strings stretching between $U(M)$ may produce massless chiral
states $(N,M,1)$, but then there is necessarily also a state $(N^*,1,M^*)$ from
the anti-brane (the notation $(*,*,*)$ refers to $U(N) \times U(M) \times U(M)_{\rm anti}$).
This cancels the $U(N)$ anomalies.
This cancellation is simply a consequence of introducing brane-antibrane pairs.
The $U(M)$ anomalies also cancel, but
for a different reason. Formula \ref{Anomaly} implies that the tadpole
multiplicities $N_a$ are such that not only the anomalies within the original
configuration cancel, but also for the CP group associated with any brane $b$
that is added to it. This is the local analog of the global anomaly
probe brane constraint, and
evidently it is automatically satisfied if the tadpoles cancel.
Note that this works in a slightly more complicated way if $A^i_{bb}$
and/or $M^i_b$ is non-zero. (Anti)-symmetric tensors contribute anomalies
$M\pm 4$. The term proportional to $M$ is cancelled by strings
in the representation $(1,M^*,M^*)$ stretching
between the brane and the anti-brane (which necessarily exist if $A^i_{bb} \not=0$),
whereas the term proportional to 4 cancels against contributions from
the probed configuration, as a consequence of (\ref{Anomaly}).

Although equations (\ref{GlobalAnomaly}) and (\ref{Anomaly}) look similar,
they are not related in any obvious way. Eqn. (\ref{Anomaly}) can be re-written
entirely in terms of left-handed fermions, but the set of labels $b$
for the two conditions is disjoint. Therefore both seem to give {\it a prori}
independent set of constraints on the Chan-Paton multiplicities $N_a$. In principle
this gives one mod-2 constraint for every symplectic boundary label $b$.
The total number of constraints is
reduced by the following considerations:
\begin{itemize}
\item{If $a$ is itself symplectic, $N_a$ is even, and hence there is no mod-2
constraint on $N_a$.}
\item{If $a$ is complex, $N_a=N_{a^c}$, which reduces the number
of independent variables.}
\item{There may be linear dependencies among the constraints.}
\item{Since the local anomaly conditions (\ref{Anomaly}) are satisfied,
so is their mod-2 reduction. Some of the global anomaly conditions may be
already contained in mod-2 reduced local anomaly conditions.}
\item{We may derive additional mod-2 constraints from the tadpole
conditions that do not produce local anomaly cancellation conditions.
This requires rewriting the remaining tadpole conditions in terms of integers,
an operation for which no canonical algorithm is known to us, while in the
previous case the anomaly takes care of that. However, all coefficients turned
out to be integers automatically in all cases we considered, after reducing the tadpole
equations to an independent set.}
\end{itemize}
Even after taking all this into account, often there still are mod-2
conditions left over, and sometimes a substantial number of them. In some
of the simpler cases, for example the tensor product $(1,1,1,1,7,16)$, there
are no global anomaly constraints at all, because there are no symplectic factors.
The next degree of complication occurs for example for
the tensor product $(1,4,4,4,4)$. It has a total of 528 MIPF/orientifold choices,
with up to 65 independent probe brane constraints. Nevertheless, in 504 cases these
are all already contained in the local anomaly conditions, and in the remaining 22 there
is just one mod-2 constraint left over. Roughly speaking, the number of left-over
mod-2 conditions increases as the tensor product contains larger tensor
factors and has more primaries. At the other extreme we have the aforementioned 
MIPF of $(1,6,46,46)$, which has 24 tadpole conditions, 10 of which 
independent from each other. From the local anomaly conditions reduced
modulo 2 we get just 2 constraints. Adding the remaining tadpole conditions
we get 10 mod-2 constraints. The symplectic factors yield 155 independent
mod-2 contraints,
which combined with the ones from the tadpoles leads to a total of
157 mod-2 constraints, and hence just 10 of these are automatically satisfied by any
solution that was previously found. Therefore the existing set of solutions
has to be checked for 147 mod-2 conditions, which could potentially
reduce the number of solutions enormously. It is very surprising that 99.7$\%$ of the
solutions survive all these constraints, as discussed above.

During the re-analysis we have used a somewhat improved method for solving
the tadpole conditions, which has allowed us to push the limits a bit furter
and solve them in a few more cases that were previously intractable.
As a result we now have more solutions than before, namely 210782, distinguished
in the same way as in \cite{Dijkstra:2004cc}. All massless spectra of this
set of solutions can be searched and examined via a webpage \cite{RCFTwebpage}.

The probe branes provide a way to define for each boundary a set of $Z_2$-charges. A priori
there can be as many charges as there are symplectic factors, but usually these are not
independent. These charges may be expected to correspond to the K-theory charges of
the corresponding D-branes in a geometric setting. We have attempted 
to make sense of these charges and tried to relate them directly to quantum
numbers of the boundary states. Unfortunately we had little success in this enterprise
except for some cases where a clear relation was found between the $q$ quantum numbers
of the boundaries\footnote{We use the standard notation $(l,q,s)$ for the quantum numbers of
the N=2 minimal models.} and the sum of the K-theory charges of the configuration.
Just as an example, for the tensor product $(1,1,2,2,4,4)$ with 110 
modular invariants, the relation holds for several orientifolds corresponding 
to 22 of these invariants (between one and four orientifolds for each invariant).

This paper leaves unanswered the important issue of a derivation, from first principles, 
of the global anomaly conditions that must be satisfied by orientifold constructions. 
Just as in other approaches, Uranga's probe brane
procedure seems to be the only method at our disposal. This is unsatisfactory and needs
to be addressed in the future, but for now the main message is that the set of solutions
is barely affected by these seemingly powerful constraints. 

\vskip .2in

\noindent
{\bf Acknowledgements:}
\vskip .2in
\noindent
We thank Mirjam Cvetic, Tim Dijkstra, Elias Kiritsis, Juan Jos\'e Manjar\'\i n, Gary Shiu, 
Angel Uranga and Ed Witten for useful conversations. This work has been partially 
supported by funding of the spanish Ministerio de Educaci\'on y Ciencia, Research Project
BFM2002-03610, and by the FOM programme "String theory and Quantum Gravity".

\bibliography{REFS}
\bibliographystyle{lennaert}

\vskip .4in

\renewcommand{\arraystretch}{1.2}
\begin{table}[h!]\caption{\footnotesize Tensor products and MIPFs for which non-trivial global anomalies affecting 
previous spectra were found. The first column specifies the tensor product,
the second the Hodge numbers of the corresponding Calabi-Yau manifold and
the number of singlets it yields in a heterotic string compactification, the
third column gives the number of boundaries, and the last a sequence number
assigned by the programme {\tt kac} \cite{KACwebpage} used to compute the spectra.}
\footnotesize
\begin{center}
~~~~~~~~~~~~~~~~\begin{tabular}{|l|c|c|c|c|c|} \hline
Tensor &  $(h_{21},h_{11},S) $ & Boundaries  & Nr.  \\ \hline \hline
         (1,6,46,46) & (9,129,525) & 1484 & 10  \\
         (1,10,22,22) & (7,55,263) & 1148 & 19 \\
                      & (20,32,237) & 1632         & 27 \\
          (2,4,14,46)       & (25,37,287) & 1152 & 8  \\
                       & (28,40,309) & 1440 & 10  \\
         (2,4,16,34)         & (26,62,339) & 1232 & 17  \\
         (2,4,22,22)      & (10,82,361) & 864 & 42  \\
                       & (13,85,367) & 1080 & 22  \\
                         & (10,58,309) & 864 & 11  \\
                       & (13,61,335) & 1080 & 13  \\
                       & (21,69,344) & 1728 & 16 \\
                         & (20,32,261) & 1668 & 17  \\
         (2,6,8,38)    & (28,52,331) & 1200 & 16  \\
                       & (22,34,265) & 720 & 25  \\
         (2,6,14,14)      & (9,57,273) & 768 & 60  \\
                       & (10,58,271) & 768 & 22  \\
                         & (9,33,233) & 768 & 21  \\
                       & (10,34,251) & 768 & 62    \\
        (2,10,10,10)        & (9,45,243) & 832 & 53    \\
                       & (13,49,251) & 1120 & 18    \\
                       & (15,51,271) & 1312 & 16    \\
                          & (19,31,231) & 1664 & 59  \\
                       & (19,31,235) & 1120 & 24  \\
          (4,4,6,22)  & (13,61,289) & 330 & 12  \\
                       & (9,33,211) & 438 & 8  \\
                       & (18,30,221) & 402 & 34  \\
\hline
\end{tabular}
\end{center}
\end{table}

\end{document}